# Large anisotropic topological Hall effect in a hexagonal non-collinear magnet $Fe_5Sn_3$

Hang Li,[1,2] Bei Ding,[1,2] Jie Chen,[1,2] Zefang Li,[1,2] Enke Liu,[1,3] Xuekui Xi,[1] Guangheng Wu,[1] and Wenhong Wang[1,3]*

[1]Beijing National Laboratory for Condensed Matter Physics, Institute of Physics, Chinese Academy of Sciences, Beijing 100190, China

[2]University of Chinese Academy of Sciences, Beijing 100049, China

[3] Songshan Lake Materials Laboratory, Dongguan, Guangdong 523808, China

*Corresponding author. Email: wenhong.wang@iphy.ac.cn

**Abstract**

We report the observation of a large anisotropic topological Hall effect (THE) in the hexagonal non-collinear magnet $Fe_5Sn_3$ single crystals. It is found that the sign of the topological Hall resistivity $\rho^{TH}$ is negative when a magnetic field H perpendicular to the bc-plane (H⊥bc-plane), however, it changes form negative to positive when H parallel to the c-axis (H∥c-axis). The value of $\rho^{TH}$ increased with the increasing temperature and reached approximately -2.12 μΩ cm (H⊥bc-plane) and 0.5 μΩ cm (H∥c-axis) at 350 K, respectively. Quantitative analyses of the measured data suggest that the observed anisotropic THE may originate from the opposite scalar spin chirality induced by the magnetic fields perpendicular and parallel to the c-axis, respectively.

The topological Hall effect (THE) as one of the spin-related effects is a hallmark of the topological chiral spin textures, such as magnetic skyrmions, which is attracting considerable interest from both fundamental and practical points of view [1,2]. In a non-collinear spin texture, the non-zero spin chirality[3,4], $\chi = S_i \cdot (S_j \times S_k)$ ($S_i$, $S_j$ and $S_k$ represent the three nearest spin) can induce non-zero Berry curvature which acts like a fictitious magnetic field for the conduction electrons and gives rise to the so-called THE[3-6]. The mechanism of THE is similar with the intrinsic anomalous Hall effect[7]. At the experimental level, the THE is found to be superposed on the ordinary and anomalous Hall effects and can be observed as an additional signal in the Hall measurements[8]. In the recent several years, the THE has served as an available electrical detection method for the topological chiral spin textures and has been widely reported in a range number of magnetic materials, such as magnetic nanostructure[5], magnetic skyrmionic materials[8-17], magnetic films[18-24], and frustrated magnetic materials[4,25,26].

For the magnetic materials used in the spintronic devices, however, the premise is that the topological chiral spin textures can be maintained at above the room temperature (RT), which motivates us to focus on the non-collinear magnets with high magnetic ordering temperatures, such as the Fe-Sn alloys family[27-30]. There are five types of alloys in the Fe-Sn compounds family, that is $Fe_3Sn$, $Fe_5Sn_3$ and $Fe_3Sn_2$, $FeSn$, and $FeSn_2$, and all of these alloys show a magnetic ordering temperature well above the RT. Among these alloys, the frustrated kagome magnet $Fe_3Sn_2$ have emerged as an important class of the magnetic topological materials in recent years because it exhibits many interesting physical properties, such as skyrmions[31-34], massive Dirac fermions[35], many-body spin-orbit tunability[36], flatbands[37], de-Haas-Van Alphen effect[38], and the large THE above the RT[16]. Furthermore, the kagome antiferromagnetic FeSn also exhibits profound relation with some interesting physical properties such as Dirac fermions and flatbands[39]. In our previous study[40], we have found a large intrinsic anomalous Hall effect in the hexagonal non-collinear magnet $Fe_5Sn_3$, which were tightly related to the non-vanishing Berry curvature due to possible non-trivial band structures. However, there is a lack of a systematic study on how the non-collinear magnetic structures further impact the magnetic and transport experiments in the $Fe_5Sn_3$ single crystals. Here, we perform the measurements of anisotropic magnetic and electrical properties in such a topologically magnetic material candidate. Interestingly, a large THE but is opposite in signal is observed when the external magnetic field is applied perpendicular to the bc-plane (H⊥bc-plane) and parallel to the c-axis (H∥c-axis) of a $Fe_5Sn_3$ single crystal. These results suggest that the observed anisotropic THE in $Fe_5Sn_3$ are associated with the opposite scalar spin chirality induced by the external magnetic field along the c-axis and ab-plane, making it a promising topological material for spintronics applications.

As shown in Fig. 1(a), the $Fe_5Sn_3$ is a hexagonal layered ferromagnet with the space group $P6_3/mmc$ (a=b=4.224 Å, c=5.222 Å). The Fe and Fe-Sn layers alternately stack along the c-axis and the three nearest atoms form the triangular meshes in every layers. In general, the triangular arrangement of the magnetic atoms tend to be form the geometrical frustrated magnetic structures in the magnetic materials[41]. The Fe atoms occupy two unequal position in the crystal structure, which are the Fe-I (0, 0, 0) and Fe-II (1/3, 2/3, 1/4), respectively. Moreover, the occupancy of Fe-II is 0.67 because there always form the vacancies or disorder at the sites of Fe-II, which is tightly related with the synthesis temperatures and further influences the $T_C$ and saturation magnetization $M_s$ in the $Fe_5Sn_3$[27,28].

Single crystals of the $Fe_5Sn_3$ were synthesized by the self-flux technique with a molar ratio of Fe : Sn =1 : 17. The mixed Fe (purity 99.95%) and Sn (purity 99.99%) grains were placed in an alumina crucible, and then the crucible was sealed into a tantalum tube under a partial Argon atmosphere. Finally, the tantalum tube was sealed in a quartz tube to avoid oxidation. The quartz tube was placed in a furnace and kept at 1,150 °C for 48 h to obtain a homogeneous metallic solution, and then the quartz tube was cooled from 910 °C to 800 °C at a rate of 1.5 °C/h and kept at 800 °C for 48 h to reduce the defects in the single crystals. In order to obtain the isolated single crystals, the quartz tube was moved quickly into the centrifuge to separate the excess Sn flux at 800 °C. The average chemical composition of the single crystals, measured by energy dispersive X-ray spectroscopy (EDS), was Fe : Sn=4.83 : 3, which was close to the value of $Fe_5Sn_3$. As shown in the inset of Fig.1 (b), the as-grown single crystal appears as a hexagonal prisms and the preferred orientation is along the c-axis of the single crystals. The plane indexes of the $Fe_5Sn_3$ single crystal, determined by the single crystal diffraction, are shown in the inset of Fig. 1(b). The pattern of X-ray diffraction, taken at room temperature by Bruker D2 x-ray machine with Cu $K_\alpha$ radiation ($\lambda$ = 1.5418 Å), is shown in Fig. 1(b), which indicates the polished rectangle surface of the samples used in this letter is (010) plane.

The magnetic and electrical measurements of the $Fe_5Sn_3$ single crystals were performed in a physical property measuring system (PPMS, Quantum Design). Figures 1(c) and (d) show the zero field cooling (ZFC) and magnetization M(H) curves (more detailed data is shown in the Fig. S1 and S2 of the Supplementary Material). The Curie temperature $T_C$ of the $Fe_5Sn_3$ is about 600 K and the easy axis is tend to be along the b-axis at low temperature, which are consistent with the former research[30,42]. Comparing the magnetization curves with H along the b-axis (H // b-axis) and c-axis (H // c-axis) at different temperatures[31], we obtained an anisotropy constant $K_u$ (see the Fig. S2 of the Supplementary Material) and the temperature dependence of the $K_u$ indicated the easy axis of the $Fe_5Sn_3$ gradually rotated from the b-axis to c-axis with the increasing temperature[31]. There was a hump of the ZFC curve near 270 K with the H⊥bc-plane, as shown in the inset of Fig. 1(c), which was similar with the situation of the hexagonal skyrmionic materials MnNiGa[11]. The magnetic data implies that the $Fe_5Sn_3$ exhibits complex non-collinear magnetic structures at the different temperatures.

The electrical measurements with H and I along different directions were measured by the standard six-probe method from 5 to 350 K, which is shown in the insets of the Figs. 2 (b), (d) and (f). In order to eliminate the influence from the voltage probe misalignment, the data of longitudinal ($\rho$) and transverse ($\rho^H$) resistivity were measured for both positive and negative field ($\rho(\mu_0 H) = [\rho(+\mu_0 H) + \rho(-\mu_0 H)]/$ 2 and $\rho^H(\mu_0 H) = [\rho^H(+\mu_0 H) - \rho^H(-\mu_0 H)]/2$). All of the curves of ρ with I along the a-axis, b-axis, and c-axis, as shown in Fig. S3 of Supplementary Material, exhibit the typical metallic behavior. The curves of magnetoresistance (MR = $[\rho(\mu_0 H) - \rho(0))/\rho(0)]$) are shown in Fig. 2(a), (c) and (e). All of their signs of MR changed from negative to positive at approximately 50 K with the decreasing temperature, which was similar to the situation of the Fe$_3$Sn$_2$[16]. All of the curves of $\rho^H$ exhibited a typical anomalous Hall effect (AHE) and the difference between with Fe$_3$Sn$_2$ is that the AHE of the Fe$_5$Sn$_3$ can keep to very low temperature. The sign change of the MR may relate to the easy axis rotation from the b-axis to c-axis with increasing temperature which also need further experimental verification, like the neutron diffraction.

In general，the formula of the Hall resistivity can be expressed as $\rho^H = R_o \mu_0 H + R_s M$, where the $R_0$ is the ordinary Hall coefficient, $\mu_0$ is the permeability of vacuum, $R_s$ is the anomalous Hall coefficient and M is the magnetization[7]. However, in some resent studies[3,4], another contribution to the Hall effect induced by the non-zero spin chirality $\chi$ in non-collinear magnetic structures has been observed. The additional Hall signal is named as the topological Hall effect. We fitted the data of $\rho^H$ with the above formula, the $R_0$ and $R_s$ can be obtained by linear fitting of the $\rho^H$ curves in high field. As shown in the insets of the Figs. 3(a)-(c), we subtracted the fitting curves (red lines) from the experimental curves (gray lines) and then obtained obvious peaks at the low field regions. The difference values between the fitting curves and experimental curves at various temperatures are shown in Figs. 3(a)-(c). In Fe$_5$Sn$_3$, as shown in the Fig. 4(b) and (d), the nearest Fe-I and Fe-II atoms form the heptamer spin clusters. In the spin clusters, the Fe-I and Fe-II bends are absent of the inversion symmetry, which induces a finite Dzyaloshinsky-Moriya interaction (DMI). The competition between the DMI and the Heisenberg exchange interaction further modulates the spin to arrange canted and induces a non-zero spin chirality in the Fe$_5$Sn$_3$, which is similar with the situation in the Fe$_{1.3}$Sb[43]. Moreover, the peaks only exist at the low field regions, which also implies that the peaks are related to the non-collinear magnetic structures in the Fe$_5$Sn$_3$ single crystals. So we think the additional contribution to the Hall effect is from the non-collinear magnetic structures in the $Fe_5Sn_3$ and hereinafter the additional Hall signal is referred to as the topological Hall resistivity $\rho^{TH}$.

As shown in Fig. 3(a)-(c), the values of $\rho^{TH}$ is negative at H⊥bc-plane and the difference between the current I along the c-axis and b-axis is very small. The absolute values of $\rho^{TH}$ at H⊥bc-plane is much larger than the values at H//c-axis. More interestingly, the signs of $\rho^{TH}$ is opposite at H⊥bc-plane and H//c-axis. The opposite signs of $\rho^{TH}$ at different direction of the $Fe_5Sn_3$ may relate to the anisotropic magnetic structures, which induces the opposite spin chirality along the c-axis and ab-plane. To the conduction electrons, the opposite spin chirality acts like an opposite fictitious magnetic field, and then induces the opposite topological Hall signal. In order to more clearly compare the values of the $\rho^{TH}$, we plotted the maximum values of the $\rho^{TH-max}$ under the different conditions, which is shown in Fig. 3(d). It is clear shown that the absolute values of $\rho^{TH-max}$ increase with the increasing temperature and the $\rho^{TH-max}$ reaches about -2.12 μΩ cm with H⊥bc-plane at 350 K which is three times large than the value for the $Fe_3Sn_2$[16].

For more clearly showing the magnetic transformation with the temperature and magnetic field in $Fe_5Sn_3$, as shown the positions of the red triangles and inverted red triangles in the insets of Figs. 3(a)-(c), we extracted the critical filed from the curves of $\rho^{TH}$(H) at various temperatures and plotted the magnetic phase diagrams for the $Fe_5Sn_3$ in the Fig 4(a) and (c). The $H_{max}$ and $H_0$ correspond to the critical fields at the maximum absolute values and the vanishing of $\rho^{TH}$, respectively. The critical fields at H//c-axis are larger than the values at H⊥bc-plane and all of the values of $H_0$ are near the saturation field ($\mu_0 H^s$) under the corresponding conditions ($\mu_0 H^s_{\|c-axis}$~1.5 $T$ and $\mu_0 H^s_{\perp bc-plane}$~1.2 $T$), which further indicates that the THE is tightly related to the non-collinear magnetic structures in the $Fe_5Sn_3$. The values of the $H_{max}$ in the $Fe_5Sn_3$ (0.3-0.5 T) is smaller than the values (0.5-0.6 T) in the $Fe_3Sn_2$, which may because the magnetic anisotropy in the $Fe_5Sn_3$ single crystal is larger than the values of $Fe_3Sn_2$. In some way, the larger anisotropy is more convenient to the formation of the topological magnetic structures[31].

In conclusion, we have measured the magnetic and electrical properties of the Fe$_5$Sn$_3$ single crystals along the different directions. For the magnetic properties, we observed the non-collinear magnetic behaviors changed with the temperatures in Fe$_5$Sn$_3$. The easy axis rotated from the b-axis to c-axis with the increasing temperature and an obvious spin reorientation signal appeared near 270 K. For the electrical properties, we found a large anisotropic topological Hall effect with the opposite sign at H⊥bc-plane and H∥c-axis, which was induced by the anisotropic non-collinear magnetic structures in the Fe$_5$Sn$_3$. The absolute values of $\rho^{TH-max}$ increased with the increasing temperature and reached approximately -2.12 μΩ cm and 0.5 μΩ cm with H⊥bc-plane and H∥c-axis at 350 K. This value of $\rho^{TH-max}$ at H⊥bc-plane is three times large than the value for Fe$_3$Sn$_2$ at 350 K. However, the detailed magnetic structures of Fe$_5$Sn$_3$ still need to further study, like using the neutron diffraction, which will improve the understanding of the relationship between the magnetic structures transformation and the electrical transport for the Fe$_5$Sn$_3$ single crystals.

See Supplementary Material for the zero field cooling (ZFC) and field cooling (FC) curves measured at H along the different directions from 5 to 400 K (Fig. S1), the magnetization curves measured at H along the different directions and the anisotropy constant K$_u$ varies with temperature (Fig. S2), the temperature dependence of the resistivity with current along the different directions at zero magnetic field (Fig. S3), the heat capacity (10-100K) and differential scanning calorimetry (DSC, 100-400K) curves (Fig. S4), the magnetoresistance (MR) along different axes of the crystals at various temperature (Fig. S5), the field dependent conductivity ($\Delta\sigma = \sigma(\mu_0 H) - \sigma(0)$) and topological hall resistivity ($\rho^{TH}$) at 200 K (Fig. S6), and the color contour mapping of the topological Hall resistivity $\rho^{TH}$ with magnetic field H and current I along different direction (Fig. S7).

This work was supported by the National Natural Science Foundation of China (No.11974406), and National Key R&D Program of China (Grant Nos.

DATA AVAILABILITY

The data that support the findings of this study are available from the corresponding author upon reasonable request.

REFERENCES

[1]N. Nagaosa and Y. Tokura, Nat. Nanotechnol **8** (12), 899 (2013).

[2]A. Fert, N. Reyren, and V. Cros, Nat. Rev. Mater. **2** (7), 17031 (2017).

[3]K. Ueda, S. Iguchi, T. Suzuki, S. Ishiwata, Y. Taguchi, and Y. Tokura, Phys. Rev. Lett. **108** (15), 156601 (2012).

[4]B. G. Ueland, C. F. Miclea, Y. Kato, O. Ayala-Valenzuela, R. D. McDonald, R. Okazaki, P. H. Tobash, M. A. Torrez, F. Ronning, R. Movshovich, Z. Fisk, E. D. Bauer, I. Martin, and J. D. Thompson, Nat. Commun. **3** (1), 1067 (2012).

[5]P. Bruno, V. K. Dugaev, and M. Taillefumier, Phys. Rev. Lett. **93** (9), 096806 (2004).

[6]G. Metalidis and P. Bruno, Phys. Rev. B **74** (4), 045327 (2006).

[7]N. Nagaosa, J. Sinova, S.Onoda, A. H. MacDonald, and N. P. Ong, Rev. Mod. Phys. **82** (2), 1539 (2010).

[8]N. Kanazawa, Y. Onose, T. Arima, D. Okuyama, K. Ohoyama, S. Wakimoto, K. Kakurai, S. Ishiwata, and Y. Tokura, Phys. Rev. Lett. **106** (15), 156603 (2011).

[9]A. Neubauer, C. Pfleiderer, B. Binz, A. Rosch, R. Ritz, P. G. Niklowitz, and P. Boni, Phys. Rev. Lett. **102** (18), 186602 (2009).

[10]T. Yokouchi, N. Kanazawa, A. Tsukazaki, Y. Kozuka, M. Kawasaki, M. Ichikawa, F. Kagawa, and Y. Tokura, Phys. Rev. B **89** (6), 064416 (2014).

[11]W. Wang, Y. Zhang, G. Xu, L. Peng, B. Ding, Y. Wang, Z. Hou, X. Zhang, X. Li, E. Liu, S. Wang, J. Cai, F. Wang, J. Li, F. Hu, G. Wu, B. Shen, and X. X. Zhang, Adv. Mater. **28** (32), 6887 (2016).

[12]Z. H. Liu, Y. J. Zhang, G. D. Liu, B. Ding, E. K. Liu, H. M. Jafri, Z. P. Hou, W. H. Wang, X.

Q. Ma, and G. H. Wu, Sci. Rep. **7** (1), 515 (2017).

[13]B. Ding, Y. Li, G. Xu, Y. Wang, Z. Hou, E. Liu, Z. Liu, G. Wu, and W. Wang, Appl. Phys. Lett. **110** (9), 092404 (2017).

[14]M. Leroux, M. J. Stolt, S. Jin, D. V. Pete, C. Reichhardt, and B. Maiorov, Sci. Rep. **8** (1), 15510 (2018).

[15]Y. Li, B. Ding, X. Wang, H. Zhang, W. Wang, and Z. Liu, Appl. Phys. Lett. **113** (6), 062406 (2018).

[16]H. Li, B. Ding, J. Chen, Z. Li, Z. Hou, E. Liu, H. Zhang, X. Xi, G. Wu, and W. Wang, Appl. Phys. Lett. **114** (19), 192408 (2019).

[17]Q. Wang, Q.Yin, and H. Lei, Chinese Phys.B **29** (1), 017101 (2020).

[18]Y. Ohuchi, Y. Kozuka, M. Uchida, K. Ueno, A. Tsukazaki, and M. Kawasaki, Phys. Rev. B **91** (24), 245115 (2015).

[19]J. Matsuno, N. Ogawa, K. Yasuda, F. Kagawa, W. Koshibae, N. Nagaosa, Y. Tokura, and M. Kawasaki, Sci. Adv. **2** (7), e1600304 (2016).

[20]K. Yasuda, R. Wakatsuki, T. Morimoto, R. Yoshimi, A. Tsukazaki, K. S. Takahashi, M. Ezawa, M. Kawasaki, N. Nagaosa, and Y. Tokura, Nat. Phys. **12** (6), 555 (2016).

[21]Y. Ohuchi, J. Matsuno, N. Ogawa, Y. Kozuka, M. Uchida, Y. Tokura, and M. Kawasaki, Nat. Commun. **9** (1), 213 (2018).

[22]C. S. Spencer, J. Gayles, N. A. Porter, S. Sugimoto, Z. Aslam, C. J. Kinane, T. R. Charlton, F. Freimuth, S. Chadov, S. Langridge, J. Sinova, C. Felser, S. Blügel, Y. Mokrousov, and C. H. Marrows, Phys.Rev. B **97** (21), 214406 (2018).

[23]L. Vistoli, W. Wang, A. Sander, Q. Zhu, B. Casals, R. Cichelero, A. Barthélémy, S. Fusil, G. Herranz, S. Valencia, R. Abrudan, E. Weschke, K. Nakazawa, H. Kohno, J. Santamaria, W. Wu, V. Garcia, and M. Bibes, Nat. Phys. **15** (1), 67 (2018).

[24]Y. Cheng, S. Yu, M. Zhu, J. Hwang, and F. Yang, Phys. Rev. Lett. **123** (23), 237206 (2019).

[25]Pradeep K. Rout, P. V. Prakash Madduri, Subhendu K. Manna, and Ajaya K. Nayak, Phys. Rev. B **99** (9), 094430 (2019).

[26]Y. Wang, J. Yan, J. Li, S. Wang, M. Song, J. Song, Z. Li, K. Chen, Y. Qin, L. Ling, H. Du, L. Cao, X. Luo, Y. Xiong, and Y. Sun, Phys. Rev. B **100** (2), 024434 (2019).

[27]H. Yamamoto, J. Phys. Soc. Jpn. **21** (6), 1058 (1966).

[28]G. Trumpy, E. Both, C. Djéga-Mariadassou, and P. Lecocq, Phys. Rev. B **2** (9), 3477 (1970).

[29]B. Fayyazi, K. P. Skokov, T. Faske, D. Yu Karpenkov, W. Donner, and O. Gutfleisch, Acta Mater. **141**, 434 (2017).

[30]B. Fayyazi, K. P. Skokov, T. Faske, I. Opahle, M. Duerrschnabel, T. Helbig, I. Soldatov, U. Rohrmann, L. Molina-Luna, K. Güth, H. Zhang, W. Donner, R. Schäfer, and O. Gutfleisch, Acta Mater. **180**, 126 (2019).

[31]Z. Hou, W. Ren, B. Ding, G. Xu, Y. Wang, B. Yang, Q. Zhang, Y. Zhang, E. Liu, F. Xu, W. Wang, G. Wu, X. Zhang, B. Shen, and Z. Zhang, Adv. Mater. **29** (29), 1701144 (2017).

[32]Z. Hou, Q. Zhang, G. Xu, C. Gong, B. Ding, Y. Wang, H. Li, E. Liu, F. Xu, H. Zhang, Y. Yao, G. Wu, X. X. Zhang, and W. Wang, Nano Lett. **18** (2), 1274 (2018).

[33]Z. Hou, Q. Zhang, G. Xu, S. Zhang, C. Gong, B. Ding, H. Li, F. Xu, Y. Yao, E. Liu, G. Wu, X. X. Zhang, and W. Wang, ACS Nano **13** (1), 922 (2019).

[34]Z. Hou, Q. Zhang, X. Zhang, G. Xu, J. Xia, B. Ding, H. Li, S. Zhang, N. M. Batra, Pmfj Costa, E. Liu, G. Wu, M. Ezawa, X. Liu, Y. Zhou, X. Zhang, and W. Wang, Adv. Mater. **32** (1), 1904815 (2020).

[35]L. Ye, M. Kang, J. Liu, F. von Cube, C. R. Wicker, T. Suzuki, C. Jozwiak, A. Bostwick, E. Rotenberg, D. C. Bell, L. Fu, R. Comin, and J. G. Checkelsky, Nature **555** (7698), 638 (2018).

[36]J. X. Yin, S. S. Zhang, H. Li, K. Jiang, G. Chang, B. Zhang, B. Lian, C. Xiang, I. Belopolski, H. Zheng, T. A. Cochran, S. Y. Xu, G. Bian, K. Liu, T. R. Chang, H. Lin, Z. Y. Lu, Z. Wang, S. Jia, W. Wang, and M. Z. Hasan, Nature **562** (7725), 91 (2018).

[37]Z. Lin, J. H. Choi, Q. Zhang, W. Qin, S. Yi, P. Wang, L. Li, Y. Wang, H. Zhang, Z. Sun, L. Wei, S. Zhang, T. Guo, Q. Lu, J. H. Cho, C. Zeng, and Z. Zhang, Phys. Rev. Lett. **121** (9), 096401 (2018).

[38]L. Ye, M. K. Chan, R. D. McDonald, D. Graf, M. Kang, J. Liu, T. Suzuki, R. Comin, L. Fu, and J. G. Checkelsky, Nat. Commun. **10** (1), 4870 (2019).

[39]M. Kang, L. Ye, S. Fang, J. S. You, A. Levitan, M. Han, J. I. Facio, C. Jozwiak, A. Bostwick, E. Rotenberg, M. K. Chan, R. D. McDonald, D. Graf, K. Kaznatcheev, E. Vescovo, D. C. Bell, E. Kaxiras, J. van den Brink, M. Richter, M. Prasad Ghimire, J. G. Checkelsky, and R. Comin,

Nat. Mater. **19** (2), 163 (2020).

[40]H. Li, B.Zhang, J. Liang, B. Ding, J. Chen, J. Shen, Z. Li, E. Liu, X. Xi, G. Wu, Y. Yao, H. Yang, and W. Wang, Phys.Rev. B **101** (14), 140409 (2020).

[41]A. P. Ramirez, Annu. Rev. Mater. Sci. **24** (1), 453 (1994).

[42]B. C. Sales, B. Saparov, M. A. McGuire, D. J. Singh, and D. S. Parker, Sci. Rep. **4** (1), 7024 (2014).

[43]Y. Shiomi, M. Mochizuki, Y. Kaneko, and Y. Tokura, Phys. Rev. Lett. **108** (5), 056601 (2012).

**Figure captions**

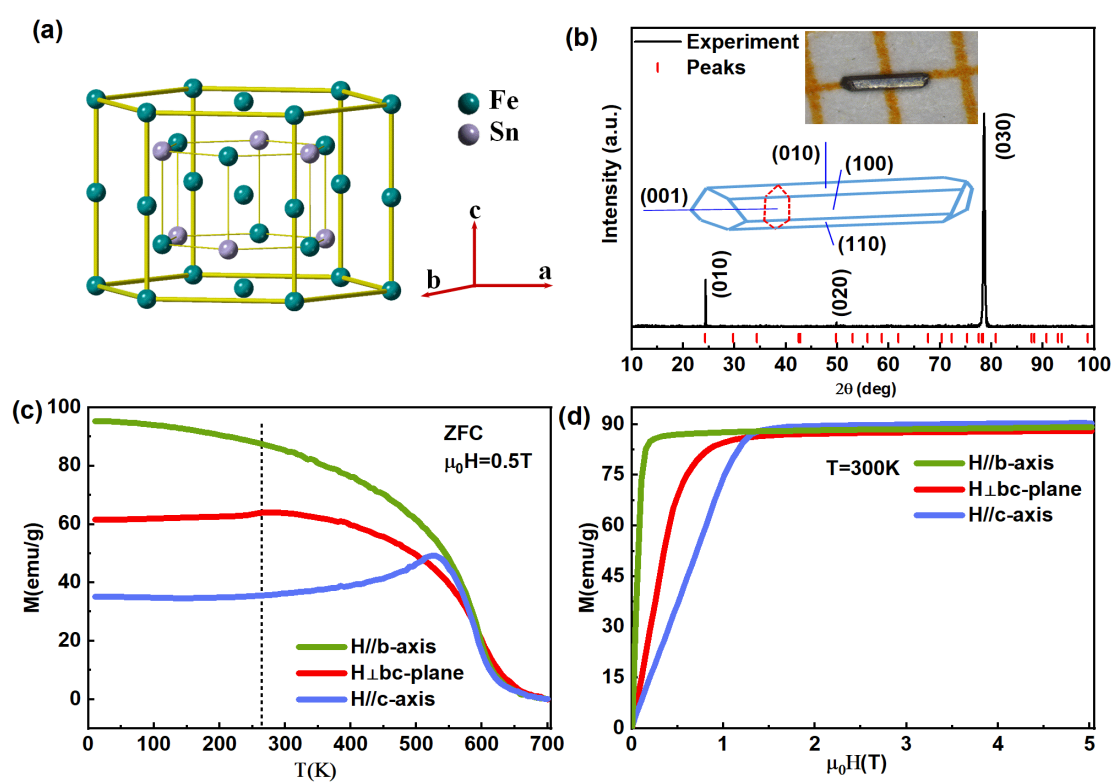

**FIG. 1** (color online). (a) The schematic of the $Fe_5Sn_3$ crystal structure. The Fe layers and Fe-Sn layers stack along c-axis alternately. (b) X-ray diffraction pattern of the $Fe_5Sn_3$ single crystal. The insets show the photo and corresponding schematic of the single crystal. (c) The ZFC curves with H along different directions of the single crystal at 0.5 T. (d) The magnetization curves with H along different directions at 300 K.

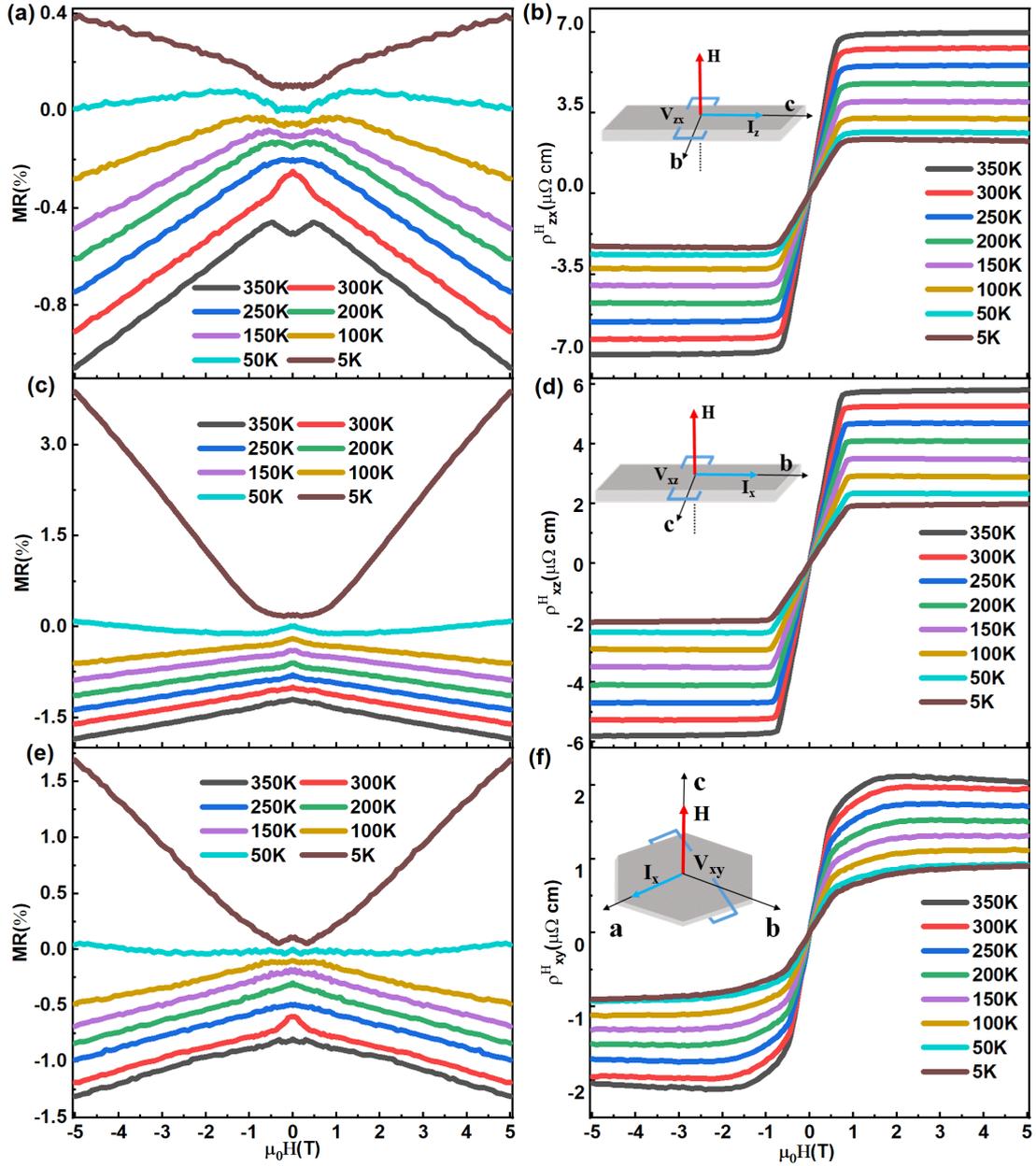

**FIG. 2** (color online). Magnetoresistance (MR) and topological Hall resistivity ($\rho^H$) with a magnetic field H and current I along different directions of the Fe$_5$Sn$_3$ single crystal. Insets of the (b), (d), and (f) are the schematic of the electrical measurement.

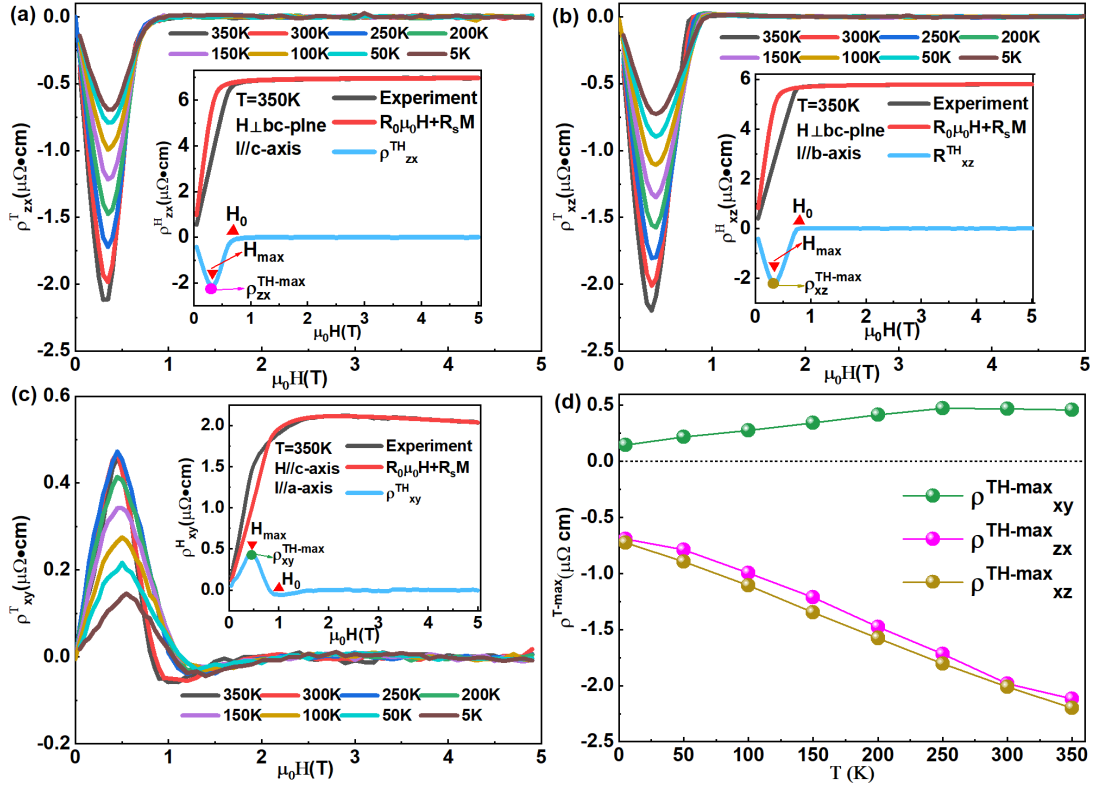

**FIG. 3** (color online). Topological Hall effect along the different directions. (a)-(c) The derived topological Hall with H and I along different directions. The gray lines represent the experiment data of Hall resistivity $\rho^H$, the red lines represent the fitting lines of the formula $R_o\mu_0 H + R_s M$ and the blue lines represent the difference between the experiment data and the fitting lines. The insets show the processes of extraction. The red inverted triangles and triangles locate at the position of the critical filed corresponding to the maximum absolute values ($H_{max}$) and vanishing ($H_0$) of the THE. (d) The maximum values of topological Hall resistivity from 5 to 350 K with H and I along the different directions.
13

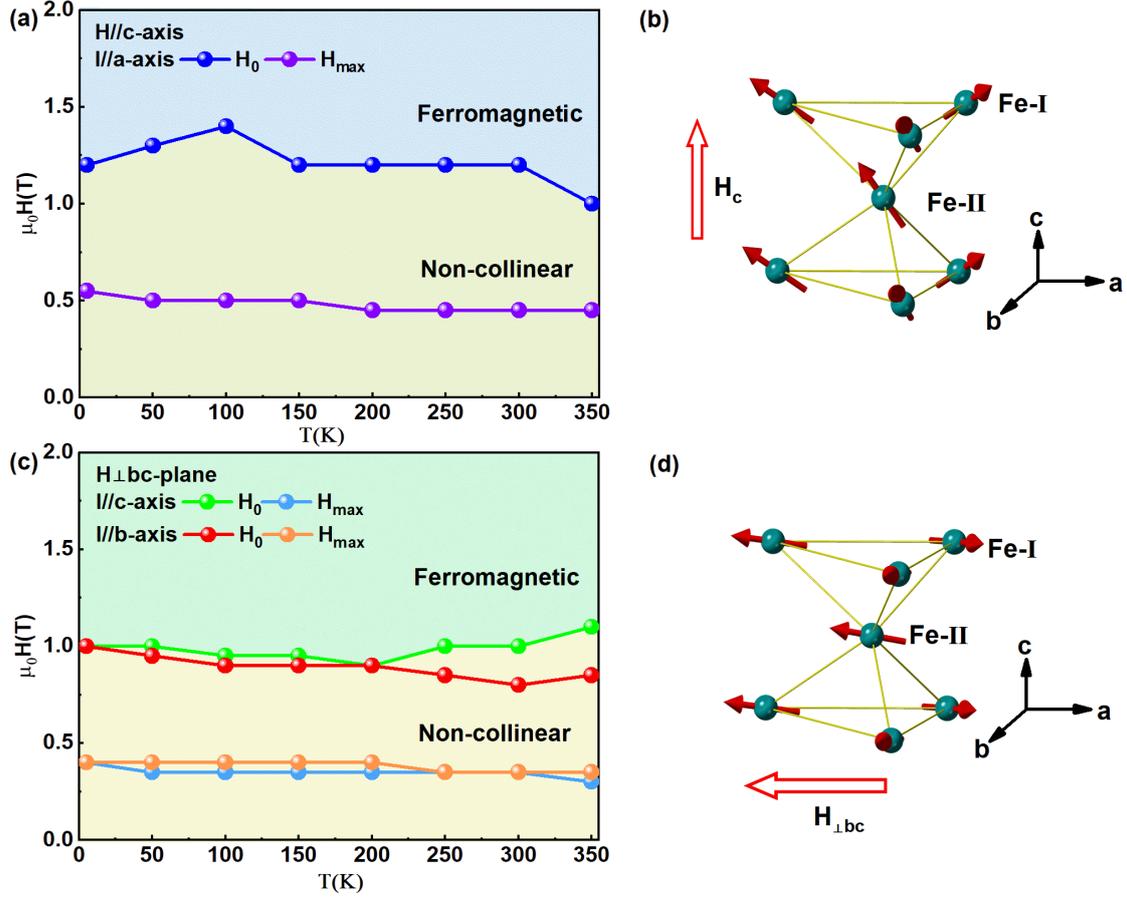

**FIG. 4** (color online). (a), (c) Magnetic phase diagram of the $Fe_5Sn_3$ single crystal with H parallel to the c-axis and perpendicular to the bc-plane, respectively. These solid circles in figures are the critical fields deduced from the topological Hall effect corresponding to the maximum absolute values ($H_{max}$) and the vanishing ($H_0$) of the topological Hall resistivities. (b), (d) The schematic of the spin structure at the external field H along different directions of the $Fe_5Sn_3$ single crystal.

# Supplementary Material for

# Large anisotropic topological Hall effect in a hexagonal non-collinear magnet $Fe_5Sn_3$

Hang Li,[1,2] Bei Ding,[1,2] Jie Chen,[1,2] Zefang Li,[1,2] Enke Liu,[1,3] Xuekui Xi[1], Guangheng Wu,[1] and Wenhong Wang[1,3*]

[1]Beijing National Laboratory for Condensed Matter Physics, Institute of Physics, Chinese Academy of Sciences, Beijing 100190, China

[2]University of Chinese Academy of Sciences, Beijing 100049, China

[3] Songshan Lake Materials Laboratory, Dongguan, Guangdong 523808, China

*Corresponding author. Email: wenhong.wang@iphy.ac.cn

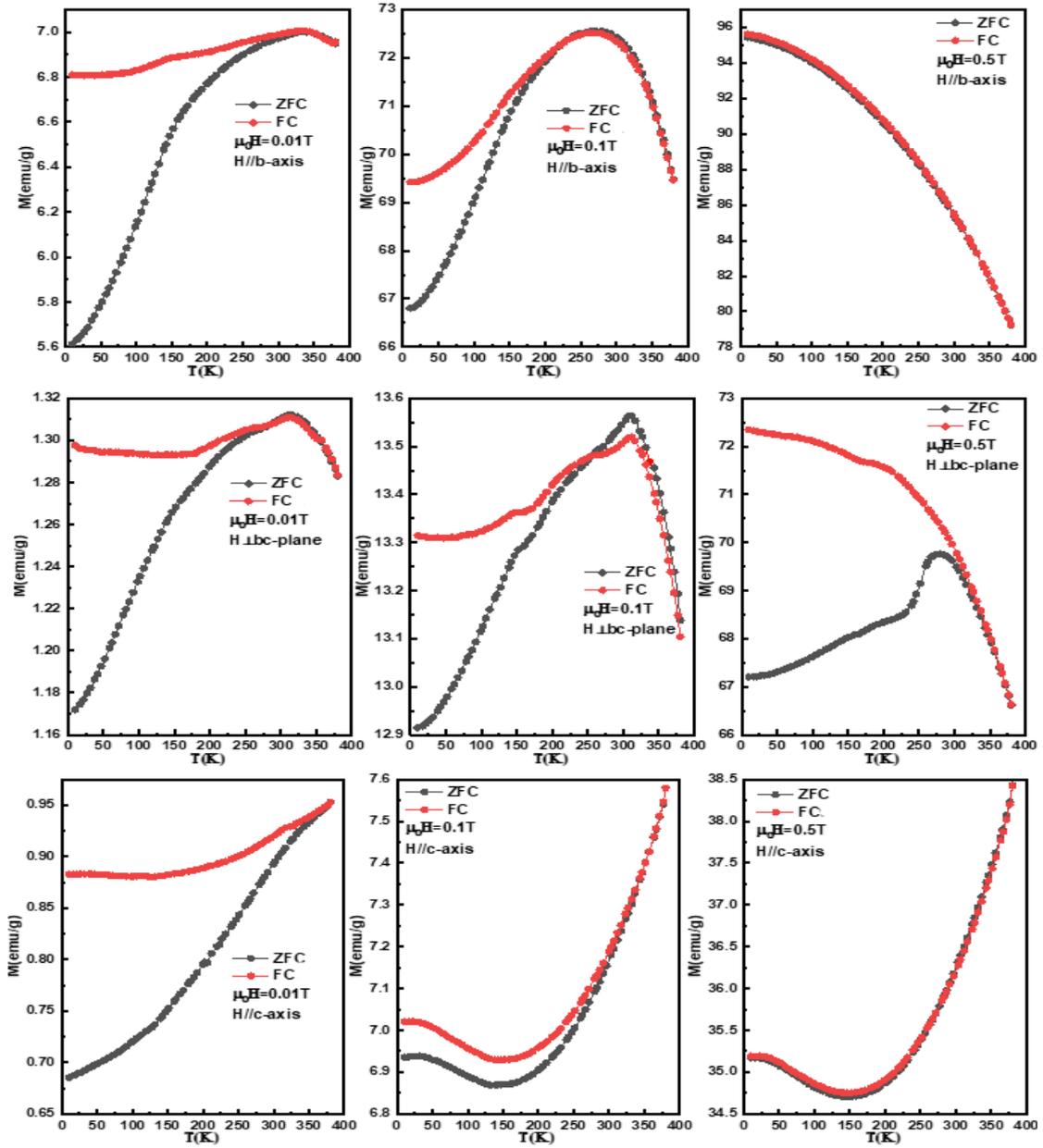

**Fig. S1** Zero field cooling (ZFC) and field cooling (FC) curves measured at H along the different directions from 5 to 400 K.

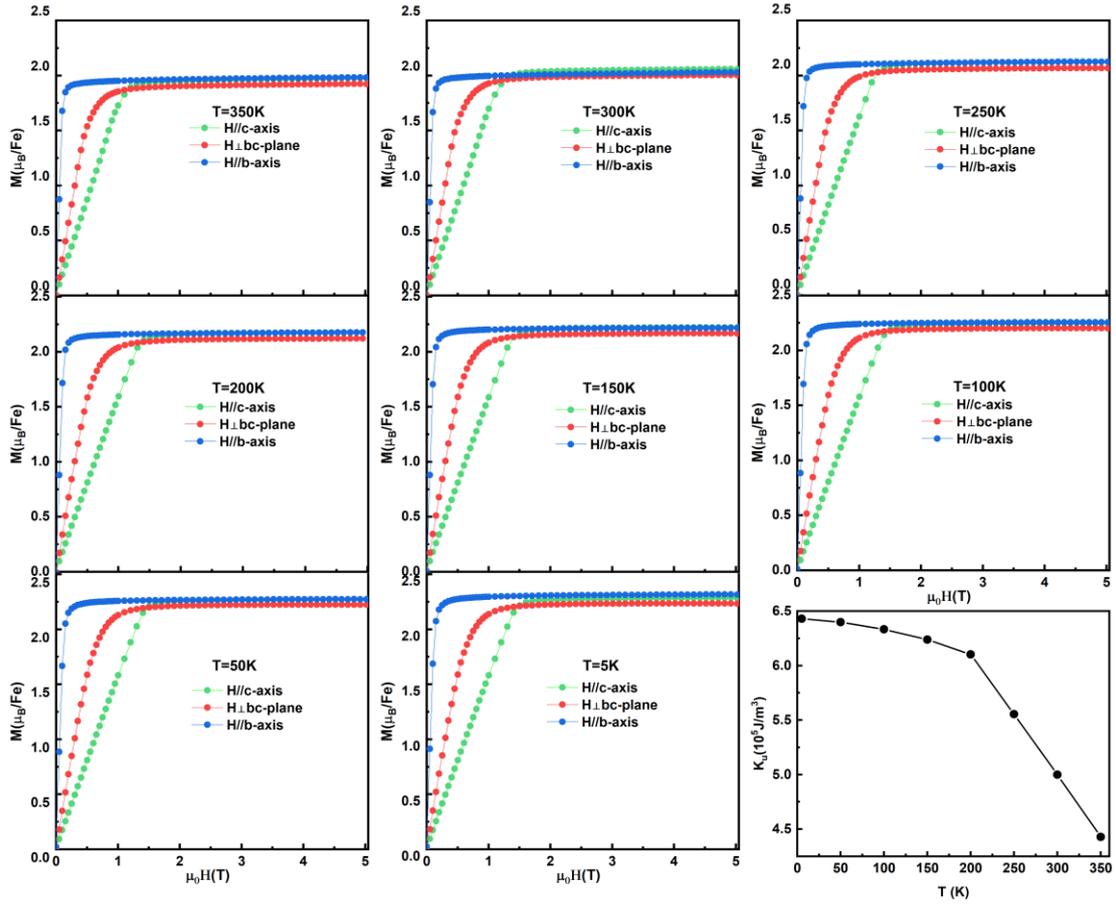

**Fig. S2** Magnetization curves measured at H along the different directions and the temperature dependent of the anisotropy constant $K_u$. The data indicates that the b-axis always more easier magnetize than the c-axis from 5 to 350 K. The anisotropy constant $K_u$ decrease with increasing temperature, which suggests that the easy magnetization direction gradually rotates to the c axis with increasing temperature.

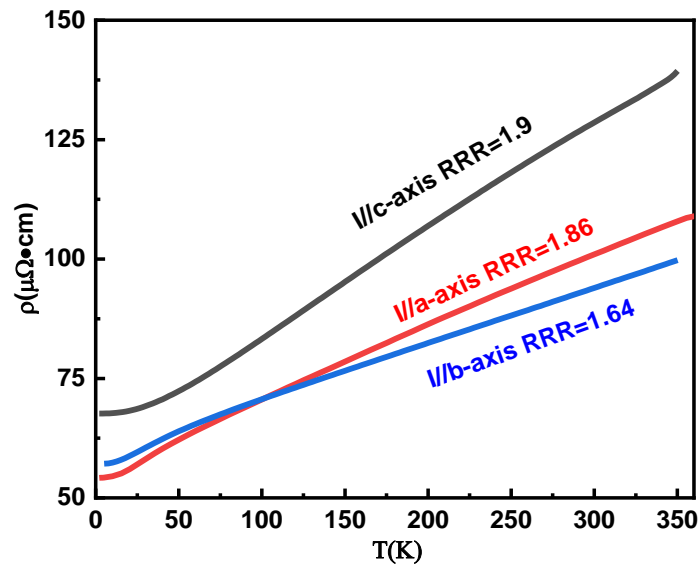

**Fig. S3** Temperature dependence of the resistivity with current along the different directions at zero magnetic field.

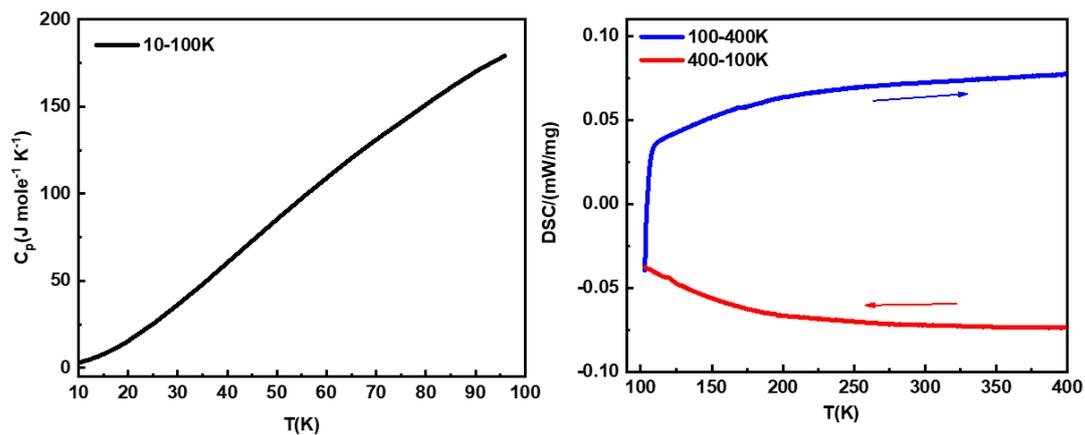

**Fig. S4** Heat capacity (10-100K) and differential scanning calorimetry (DSC, 100-400K) curves. Both of them didn't show any signals about the crystal structure change with temperature.

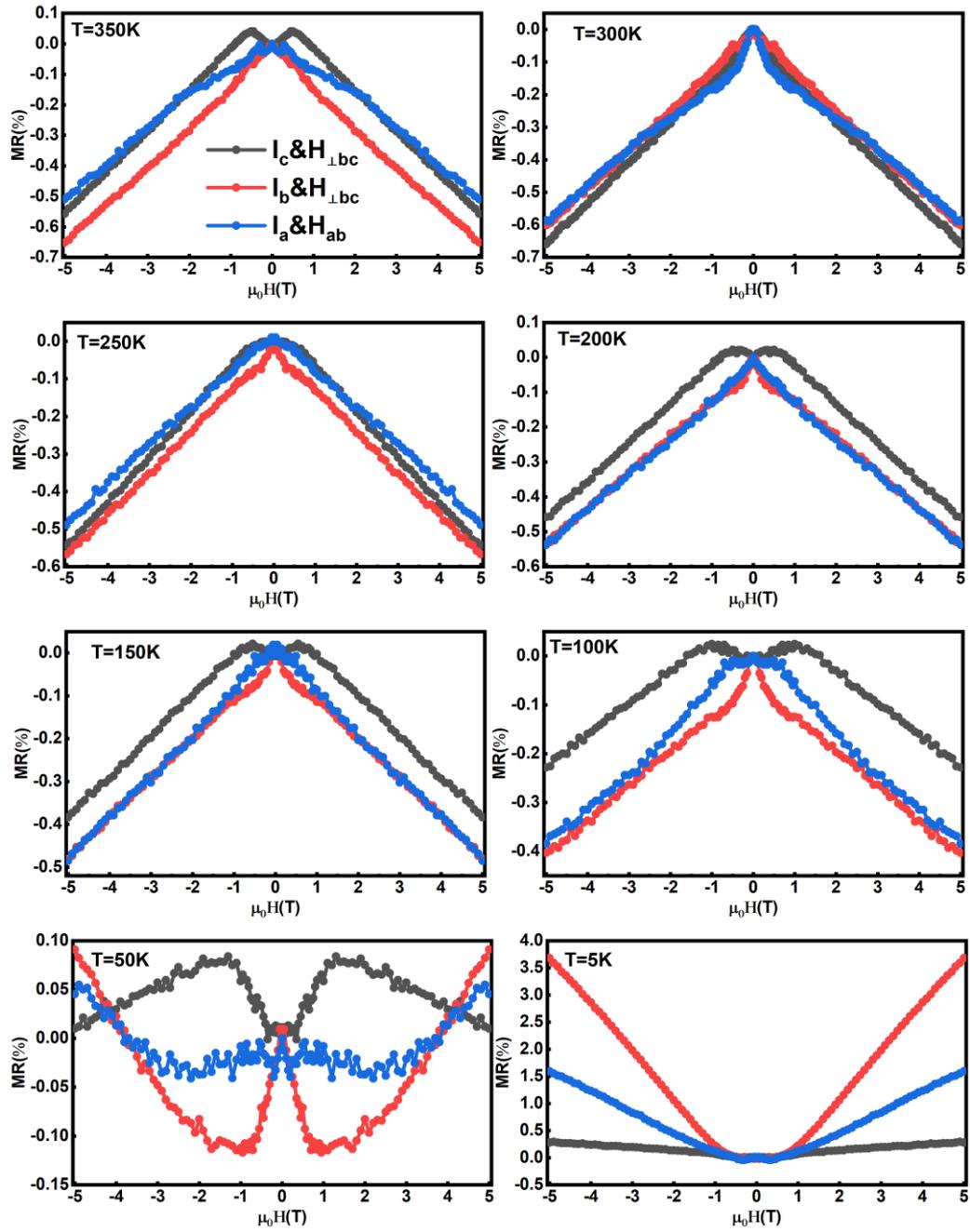

**Fig. S5** Magnetoresistance (MR) along different axes of the crystals at various temperature.

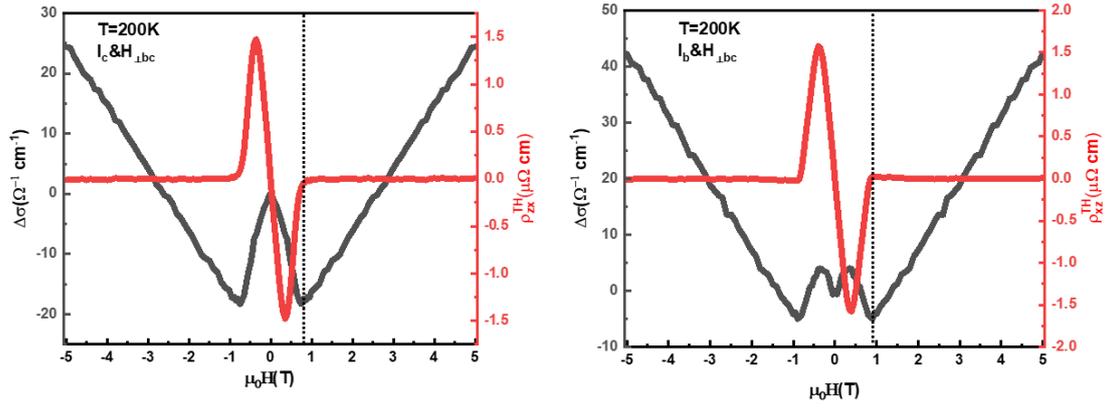

**Fig. S6** Field dependent conductivity ($\Delta\sigma = \sigma(\mu_0 H) - \sigma(0)$) and topological hall resistivity ($\rho^{TH}$) at 200 K. Their kink points is consistent with each other which suggests that the changes of MR at low field zone are related to the metamagnetic transition, where the magnetic structure change with the extrinsic magnetic field. The situation at other temperatures are similar with this.

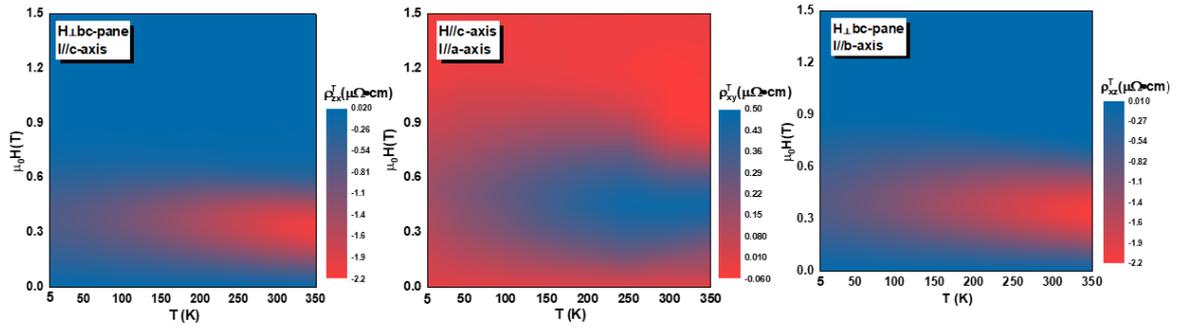

**Fig. S7** Color contour mapping of the topological Hall resistivity $\rho^{TH}$ with magnetic field H and current I along different directions.